\documentclass[letterpaper]{article} 
\usepackage{aaai2026}  
\usepackage{times}  
\usepackage{helvet}  
\usepackage{courier}  
\usepackage[hyphens]{url}  
\usepackage{graphicx} 
\usepackage{natbib}  
\usepackage{caption} 
\usepackage{algorithm}
\usepackage{algorithmic}

\usepackage{verbatim}
\usepackage{multirow}
\usepackage{lettrine}
\usepackage[utf8]{inputenc}
\usepackage{newunicodechar}
\newunicodechar{，}{,}
\usepackage{textcomp}
\usepackage{booktabs}
\usepackage{bbding}
\usepackage{pifont}
\newcommand{\cmark}{\ding{51}}
\newcommand{\xmark}{\ding{55}}

\usepackage{bm}
\usepackage{xcolor}
\usepackage{colortbl}
\usepackage{subcaption}
 
\usepackage{amsmath}
\usepackage{amsfonts}
\usepackage{amssymb}

\usepackage{newfloat}
\usepackage{listings}
\DeclareCaptionStyle{ruled}{labelfont=normalfont,labelsep=colon,strut=off} 
\floatstyle{ruled}
\newfloat{listing}{tb}{lst}{}
\floatname{listing}{Listing}
\title{WaveC2R: Wavelet-Driven Coarse-to-Refined Hierarchical Learning \\for Radar Retrieval}
\author{
    Chunlei Shi\textsuperscript{\rm 1},
    Han Xu\textsuperscript{\rm 1},
    Yinghao Li\textsuperscript{\rm 2},
    Yi-Lin Wei\textsuperscript{\rm 2},
    Yongchao Feng\textsuperscript{\rm 3},\\
    Yecheng Zhang\textsuperscript{\rm 4},
    Dan Niu\textsuperscript{\rm 1}\thanks{Corresponding author.}
}

\affiliations{
    \textsuperscript{\rm 1}School of Automation, Southeast University\\
    \textsuperscript{\rm 2}School of Computer Science and Engineering, Sun Yat-sen University\\
    \textsuperscript{\rm 3}State Key Laboratory of Virtual Reality Technology and Systems, Beihang University\\
    \textsuperscript{\rm 4}School of Architecture, Tsinghua University\\
    danniu1@163.com
}

\begin{document}

\maketitle

\begin{abstract}
Satellite-based radar retrieval methods are widely employed to fill coverage gaps in ground-based radar systems, especially in remote areas affected by terrain blockage and limited detection range. Existing methods predominantly rely on overly simplistic spatial-domain architectures constructed from a single data source, limiting their ability to accurately capture complex precipitation patterns and sharply defined meteorological boundaries.
To address these limitations, we propose WaveC2R, a novel wavelet-driven coarse-to-refined framework for radar retrieval. WaveC2R integrates complementary multi-source data and leverages frequency-domain decomposition to separately model low-frequency components for capturing precipitation patterns and high-frequency components for delineating sharply defined meteorological boundaries. Specifically, WaveC2R consists of two stages \textit{(i)} \textit{Intensity-Boundary Decoupled Learning}, which leverages wavelet decomposition and frequency-specific loss functions to separately optimize low-frequency intensity and high-frequency boundaries; and \textit{(ii)} \textit{Detail-Enhanced Diffusion Refinement}, which employs frequency-aware conditional priors and multi-source data to progressively enhance fine-scale precipitation structures while preserving coarse-scale meteorological consistency.
Experimental results on the publicly available SEVIR dataset demonstrate that WaveC2R achieves state-of-the-art performance in satellite-based radar retrieval, particularly excelling at preserving high-intensity precipitation features and sharply defined meteorological boundaries. 
\end{abstract}

\begin{links}
    \link{Project}{{https://spring-lovely.github.io/WaveC2R/}}
\end{links}

\begin{figure}[ht]
\begin{center}
\centerline{\includegraphics[width=0.95\columnwidth]{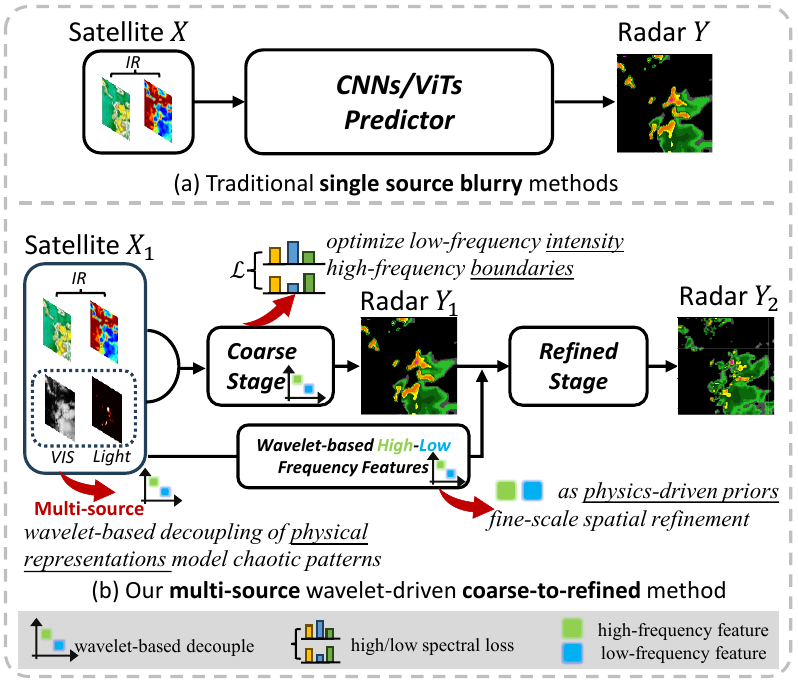}}
\caption{Comparison between traditional and proposed radar retrieval approaches. (a) Traditional methods employ spatial domain architectures with single-source infrared data (IR). (b) Our WaveC2R integrates multi-source satellite observations (visible, infrared, and lightning) through coarse-to-refined stages: initial estimation followed by conditional diffusion refinement with physics-driven priors.}
\label{fig:motivation}
\end{center}
\vskip -0.39in
\end{figure}
\section{Introduction}
\label{intro}

Satellite-based radar retrieval plays a critical role in filling coverage gaps of ground-based radar systems, providing essential precipitation monitoring in data-sparse regions \cite{park2025data, wen2024duocast}.
Geostationary satellites such as Himawari-8, GOES-R, and Fengyun-4A provide high-resolution meteorological observations \cite{liu2020noaa, jing2022cascaded} for radar composite reflectivity reconstruction.
These satellites capture visible, infrared, and lightning data, which collectively enhance characterization of complex precipitation patterns in regions lacking ground-based radar coverage \cite{li2023toward, yang2023multitask}.

Traditional retrieval methods, such as radiative transfer models (RTMs) \cite{di2018enhancing} and variational methods \cite{yin2021impact}, rely heavily on physical models and statistical techniques. These methods are computationally intensive and often constrained in accuracy due to various factors. In recent years, deep learning methods have become a hot topic in radar retrieval research due to their powerful data processing capabilities and efficient predictive performance \cite{jiang2024retrieving, park2025data}. However, despite the notable progress in deep learning-based satellite radar retrieval, current methods still suffer from several critical limitations that restrict their performance (Fig.\ref{fig:motivation}). 

\textbf{Single-source data limitations.} Most current approaches rely solely on single-source satellite data, typically infrared channels. This neglects complementary information from alternative data sources. Visible observations provide crucial cloud-top morphological details during daylight hours that enhance structural identification. Similarly, lightning data serves as a direct indicator of convective intensity and enables precise storm core localization \cite{yu2025integrating, zheng2024cross, zhao2024weathergfm, chen2024terra}. Without these multi-source inputs, current methods struggle with incomplete scene representation, limiting their ability to accurately characterize severe convective cells \cite{jin2023rda, niu2024fsrgan}. 

\textbf{Frequency domain optimization challenges.} Accurate characterization of convective-scale structures requires effective modeling of meteorological systems' multi-frequency characteristics \cite{jin2023rda}. However, existing satellite-based radar retrieval methods predominantly employ spatial domain optimization with simplistic architectures and L2 loss functions \cite{zhao2024intelligent}. This approach fails to capture the inherent frequency characteristics of weather phenomena, resulting in oversmoothed gradients, blurred precipitation boundaries, and underestimated high-intensity convective cores \cite{yang2023deep, hilburn2020development}.

\textbf{Limited diffusion model exploration in radar retrival.} Diffusion models offer significant potential for satellite-based radar retrieval through their ability to model complex probability distributions and progressively refine meteorological structures. Despite proven success in various image generation tasks, diffusion models remain unexplored in satellite-based radar retrieval. Traditional approaches struggle with recovering high-frequency meteorological details, particularly in convective regions with steep gradients \cite{gong2024cascast, wen2024duocast, li2024scrd}. 

To address these fundamental limitations, we propose WaveC2R, a novel \underline{wave}let-driven \underline{c}oarse-\underline{t}o-\underline{r}efined hierarchical learning framework for satellite-to-radar retrieval. Our approach first integrates complementary multi-source satellite observations (VIS, IR, Light) through hierarchical frequency-domain feature decoupling and fusion, enabling comprehensive capture of complex meteorological patterns across temporal-frequency dual domains. Based on our key insight that low-frequency wavelet components primarily encode precipitation intensity distribution while high-frequency components capture sharp meteorological boundaries (see Fig.~\ref{fig:wave}), WaveC2R decomposes the radar retrieval task into separate optimization processes for intensity prediction and boundary localization. Specifically, WaveC2R consists of two stages: the coarse stage employs a Wavelet-Temporal-Frequency (WTF) module with our Frequency-decomposed Intensity-Boundary Loss (FIBL) that separately constrains low-frequency intensity patterns and high-frequency boundary components, while the refined stage pioneers conditional diffusion models with physics-aware frequency-decomposed priors for progressive detail enhancement. By explicitly modeling the distinct physical roles of wavelet frequency components, WaveC2R provides more accurate radar retrieval than existing methods, particularly excelling in preserving high-intensity precipitation structures and sharp meteorological boundaries. The major contributions of this paper can be summarized as:
\begin{figure}[t]
\begin{center}
\centerline{\includegraphics[width=\columnwidth]{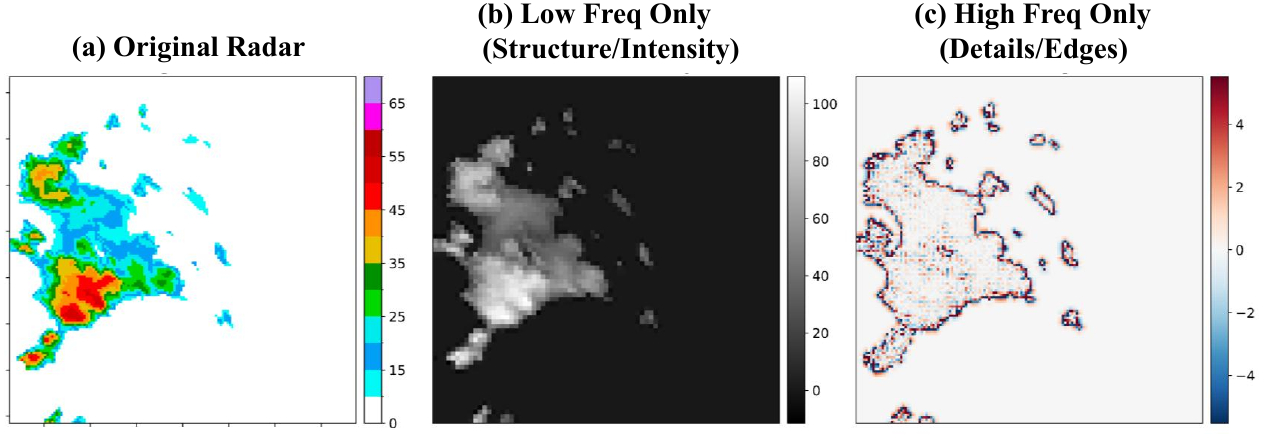}}
\caption{Wavelet-based frequency decomposition of radar reflectivity via DWT and selective IDWT reconstruction reveals distinct physical roles: (a) Original radar data, (b) Low-frequency reconstruction encodes precipitation intensity distribution, (c) High-frequency reconstruction captures boundary structures and meteorological details.}
\label{fig:wave}
\end{center}
\vskip -0.35in
\end{figure}
\begin{itemize}
    \item We propose a temporal-frequency dual-domain fusion framework that integrates complementary multi-source satellite observations and establishes physical interpretability of wavelet decomposition, revealing that low-frequency components encode intensity distribution while high-frequency components capture spatial boundaries for meteorological structure modeling.
    
    \item We introduce the wavelet-temporal-frequency module that performs hierarchical frequency-domain decomposition with our frequency-decomposed intensity-boundary loss, which disentangles intensity patterns from boundary components and mitigates energy imbalance inherent in conventional spatial-domain optimization.
    
    \item We design a physics-aware conditional diffusion framework for satellite-based radar retrieval that leverages frequency-decomposed priors to progressively refine meteorological details while preserving coarse-scale structure coherence through counterfactual feature enhancement.
\end{itemize}

\section{Related Work}
\begin{figure*}[t]
\centering
\includegraphics[width=\textwidth]{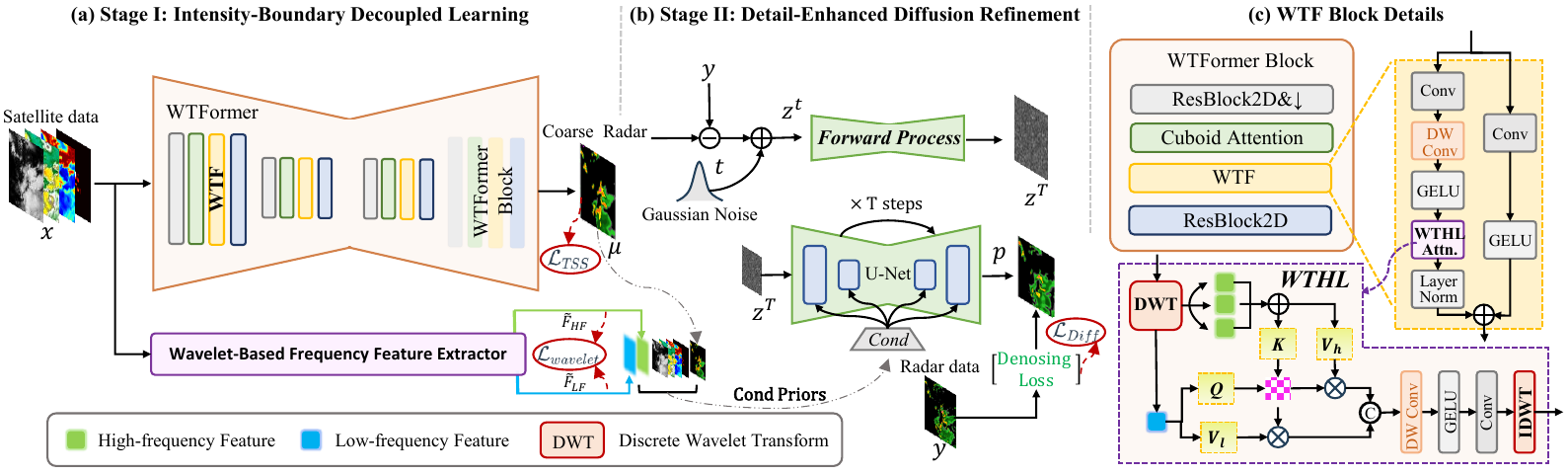}
\caption{Overall architecture of the proposed WaveC2R framework. (a) \textit{Stage I: Intensity-Boundary Decoupled Learning} employs the Wavelet-Temporal-Frequency (WTF) module to extract hierarchical frequency-domain features from multi-source satellite observations and generates coarse radar estimates through frequency-decomposed optimization. (b) \textit{Stage II: Detail-Enhanced Diffusion Refinement} progressively refines coarse estimates via conditional diffusion with physics-aware frequency-decomposed priors. (c) \textit{WTF Block} demonstrates the core wavelet-based attention mechanism that performs temporal-frequency feature fusion across multiple scales.}
\label{fig:framework}
\end{figure*}
\subsection{Radar Retrieval Using Satellite Observations}

Deep learning has emerged as a transformative approach for radar composite reflectivity (CREF) reconstruction by leveraging satellite observations. UNet-based models have become predominant due to their encoder-decoder structure and multi-scale feature extraction capabilities, with substantial improvements demonstrated through multi-band satellite integration \cite{jin2023rda, hilburn2020development, zhao2024intelligent}. Alternative architectural innovations have emerged to address UNet limitations, including residual deep forest models \cite{bao2024infrared} and integrated frameworks combining multi-band infrared data \cite{wang2020infrared}. Recent investigations have highlighted the importance of integrating diverse satellite observations encompassing visible, infrared, and lightning data to capture spatiotemporal information \cite{si2023novel, yu2025integrating, zheng2024cross, zhao2024weathergfm, chen2024terra}. However, existing methodologies continue to face challenges in effectively integrating multi-channel data, often relying on simple concatenation strategies that fail to exploit the complementary strengths of different satellite observations.

\subsection{Frequency-Domain and Wavelet Transform in Meteorological Applications}

Wavelet transform has established itself as a powerful technique for multi-scale meteorological analysis, offering distinctive advantages over traditional Fourier methods when processing non-stationary weather signals \cite{saoud2021wind}. The integration of frequency-domain techniques with deep learning has yielded promising results, with improvements achieved through Fast Fourier Transform integration in weather forecasting frameworks \cite{li2025climatellm} and specialized frequency-domain learning methods \cite{wang2024fredf}. The multi-resolution capabilities of wavelet transforms have proven particularly valuable for precipitation systems, with enhanced predictions resulting from combining wavelet transforms with time series models \cite{ahn2023short, mammedov2022weather} and frequency domain feature extraction fusion mechanisms \cite{liao2024wind}. However, existing wavelet-based approaches in meteorology primarily focus on mathematical decomposition without exploring the distinct physical roles of different frequency components in precipitation systems.

In this work, we approach radar retrieval from two disentangled frequency-domain perspectives: the intensity distribution (low-frequency) and spatial boundaries (high-frequency) of precipitation systems, distinguishing it from previous approaches that treat frequency components as unified mathematical decompositions.

\section{Method} 
\subsection{Problem Definition}
We formulate satellite-based radar retrieval as a cross-modal reconstruction problem. Given synchronous multi-source satellite observations $\mathbf{X} = \{X_{\text{vis}}, X_{\text{ir}}, X_{\text{light}}\}$ comprising visible, infrared (dual-channel), and lightning data with total channels $C=4$, our objective is to retrieve the corresponding radar reflectivity field $\hat{\mathbf{Y}} \in \mathbb{R}^{B \times 1 \times H \times W}$ that accurately approximates ground truth radar observations $\mathbf{Y} \in \mathbb{R}^{B \times 1 \times H \times W}$.

This retrieval problem is formulated as:
\begin{equation}
\hat{\mathbf{Y}} = f_{\theta}(\mathbf{X}), \quad \theta^* = \arg\min_{\theta} \mathbb{E}[\mathcal{L}(f_{\theta}(\mathbf{X}), \mathbf{Y})],
\end{equation}
where $f_{\theta}$ represents a learnable retrieval mapping function, and $\mathcal{L}$ is a physics-informed loss incorporating meteorological constraints.

\subsection{Preliminaries: Wavelet Transform}
Given a 2D feature $\mathbf{F} \in \mathbb{R}^{H \times W}$, the discrete wavelet transform (DWT) decomposes it into four frequency sub-bands through separable 1D transforms applied along rows and columns: 
\begin{equation}
\mathbf{F}_{ll}, \mathbf{F}_{lh}, \mathbf{F}_{hl}, \mathbf{F}_{hh} = \text{DWT}(\mathbf{F}),
\end{equation}
where $\mathbf{F}_{ll} \in \mathbb{R}^{H/2 \times W/2}$ represents the low-frequency approximation preserving primary structural features, while $\{\mathbf{F}_{lh}, \mathbf{F}_{hl}, \mathbf{F}_{hh}\}$ capture horizontal, vertical, and diagonal high-frequency details including boundaries and fine-grained meteorological structures. 

The DWT operation is implemented through convolution with wavelet basis functions $f_l = \phi$ (low-pass) and $f_h = \psi$ (high-pass):
\begin{equation}
\mathbf{F}_{ab}[h,w] = \sum_{i,j} \mathbf{F}[i,j] \cdot f_a[i-2h] \cdot f_b[j-2w], 
\end{equation}
where $a,b \in \{l,h\}$, $h \in [0, H/2-1]$ and $w \in [0, W/2-1]$ represent the spatial coordinates in the downsampled wavelet domain. The inverse transform IDWT reconstructs the original feature from all four sub-bands, providing the foundation for frequency-domain feature manipulation in meteorological applications.

\subsection{WaveC2R Framework}
We propose WaveC2R, a two-stage coarse-to-refined framework for satellite-based radar retrieval that addresses the coupled intensity-boundary optimization challenge through physics-aware frequency-domain decomposition. The architecture, illustrated in Fig.~\ref{fig:framework}, follows a sequential mapping: $\mathbf{X} \xrightarrow{\text{Stage I}} \boldsymbol{\mu} \xrightarrow{\text{Stage II}} \hat{\mathbf{Y}}$. Our key insight is that precipitation intensity distribution manifests primarily in low-frequency wavelet components, while precipitation boundaries are captured in high-frequency details, enabling separate optimization objectives for intensity prediction and boundary localization. This framework operates through two complementary stages: coarse structure establishment via frequency-domain decomposition, followed by diffusion-based detail refinement.

\noindent \textbf{Intensity-Boundary Decoupled Learning.}
We establish intensity-boundary decoupling through wavelet decomposition, recognizing that precipitation intensity and boundary localization exhibit distinct frequency characteristics. This is realized via two key components: the Wavelet-Temporal-Frequency (WTF) module for hierarchical multi-source feature fusion, and the Frequency-decomposed Intensity-Boundary Loss (FIBL) that separately constrains low-frequency intensity patterns and high-frequency boundary components. The WTF module extracts frequency-specific features from satellite inputs, then separate decoder branches independently reconstruct intensity and boundary components under FIBL supervision, generating the coarse radar estimation $\boldsymbol{\mu}$ with physically-consistent modeling (see Fig.~\ref{fig:framework}(a)).

\noindent \textbf{Detail-Enhanced Diffusion Refinement.}
To further enhance fine-scale meteorological structures while maintaining physical consistency, this refinement stage takes the coarse radar estimation $\boldsymbol{\mu}$ from Stage I and progressively refines it into the final high-fidelity prediction $\hat{\mathbf{Y}}$ through conditional diffusion modeling. The diffusion model employs frequency-aware conditional priors to progressively enhance fine-scale precipitation patterns while preserving coarse-scale meteorological consistency through iterative denoising guided by wavelet-decomposed features. This stage produces the final radar estimate with enhanced boundary precision and preserved meteorological structure (see Fig.~\ref{fig:framework}(b)).

\subsection{Stage I: Intensity-Boundary Decoupled Learning}
Existing methods jointly optimize precipitation intensity and boundary characteristics, leading to oversmoothed results in spatial domain. We discover that intensity distribution and boundary localization exhibit distinct frequency signatures: intensity manifests in low-frequency wavelet components while boundaries concentrate in high-frequency details (Fig.~\ref{fig:wave}). Based on this insight, we establish frequency-domain decoupling that separates these optimization objectives, enabling specialized learning strategies for accurate precipitation structure reconstruction.

\subsubsection{WTFormer Architecture.}
In the coarse retrieval stage, we employ a WTFormer architecture to generate the coarse radar estimate $\boldsymbol{\mu}$ from multi-source satellite observations $\mathbf{X} \in \mathbb{R}^{B \times C \times H \times W}$. Drawing inspiration from successful encoder-decoder architectures in atmospheric science applications, WTFormer adopts a hierarchical U-Net structure with two encoding levels and corresponding decoding levels. Each level strategically integrates three key components: ResBlock2D for multi-scale down/upsampling, Cuboid Attention mechanism from Earthformer for capturing spatiotemporal dependencies, and our proposed WTF module for frequency-aware enhancement.
The encoder $\mathcal{E}(\cdot)$ progressively extracts multi-scale features through these integrated components, while the symmetric decoder $\mathcal{D}(\cdot)$ reconstructs the coarse estimate via skip connections. The overall architecture can be formulated as:
\begin{equation}
\boldsymbol{\mu} = \text{WTFormer}(\mathbf{X}) = \mathcal{D}(\mathcal{E}(\mathbf{X}))
\end{equation}
To capture both intensity distribution and boundary characteristics, WTFormer generates a coarse radar estimate $\boldsymbol{\mu}$ through progressive optimization with FIBL loss. Detailed architectural specifications are provided in Appendix A.1.
\subsubsection{WTF Block.}
Given an input feature map $\mathbf{F}_{\text{in}}$, the WTF Block follows a dual-branch structure (see Fig.~\ref{fig:framework}(c)): (1) a convolutional pathway with channel expansion, depthwise convolutions, and nonlinear activations, and (2) a wavelet-based pathway that processes features through the Wavelet Transform High-Low (WTHL) attention module. 

In the WTHL module, discrete wavelet transform (DWT) decomposes input features into low-frequency approximation ($\mathbf{F}_{\text{LF}}$) and aggregated high-frequency details ($\mathbf{F}_{\text{HF}}$). These components are processed through a cross-frequency attention mechanism where low-frequency features serve as queries ($\mathbf{Q} = \mathbf{F}_{\text{LF}}\mathbf{W}_Q$) to attend to high-frequency keys ($\mathbf{K} = \mathbf{F}_{\text{HF}}\mathbf{W}_K$), computing attention weights via $\text{softmax}(\mathbf{Q}\mathbf{K}^T/\sqrt{d})$ and generating weighted outputs for both frequency domains. The refined features are then reconstructed using IDWT:
\begin{equation}
\hat{\mathbf{F}}_{\text{WTHL}} = \text{IDWT}(\text{Conv}(\text{GELU}(\mathbf{A}_h + \mathbf{A}_l)))
\end{equation}

The dual pathways are combined through residual connections, enabling the WTF Block to capture both conventional spatial features and frequency-domain interactions for enhanced multi-source satellite data fusion.
\subsubsection{FIBL and Temperature Scheduling Strategy}

Traditional spatial-domain losses treat all frequency components equally, leading to energy imbalance where low-frequency intensity patterns dominate optimization while high-frequency boundary details are under-represented. Although Fourier Global Loss (FGL)\cite{yan2024fourier} effectively captures global motion trends, it struggles with local positioning accuracy due to its global frequency spectrum operation, making it insufficient for capturing localized high-frequency structures that define precipitation boundaries.

To address these limitations, we propose FIBL that leverages wavelet decomposition to simultaneously preserve both frequency characteristics and spatial localization. Unlike FGL's global frequency constraints, FIBL employs spatially-localized wavelet coefficients that maintain the spatial relationships essential for accurate boundary preservation. FIBL decomposes the optimization objective into two complementary components:
\begin{equation}
\mathcal{L}_{\text{FIBL}} = \mathcal{L}_{\text{low}} + \alpha \cdot \mathcal{L}_{\text{high}},
\end{equation}
where, $\mathcal{L}_{\text{low}} = \text{MSE}(\hat{F}_{ll}, {F}_{ll})$, $\mathcal{L}_{\text{high}} = \sum_{d \in \{lh,hl,hh\}} w_d \cdot \text{MSE}(\hat{F}_d, {F}_d)$. $\hat{F}_{ll}, {F}_{ll}$ represent the low-frequency approximation components of prediction $\boldsymbol{\mu}$ and ground truth $\mathbf{Y}$, while $\hat{F}_d, {F}_d$ denote high-frequency details in horizontal ($lh$), vertical ($hl$), and diagonal ($hh$) directions. The parameter $\alpha$ is determined by energy distribution statistics between frequency components (Appendix B.2), and $w_d$ weights account for varying importance of directional boundary characteristics in meteorological structures.

To ensure proper optimization progression that aligns with physical storm development, we employ a temperature scheduling strategy that progressively transitions from FGL (computation details in Appendix B.2) to FIBL during training. This reflects the natural meteorological development where intensity patterns establish before boundary refinement. Using cosine annealing $P(t) = \cos(\pi t/(2T))$ and random sampling $p \sim \mathcal{U}(0,1)$, the training loss becomes:
\begin{equation}
\mathcal{L}_{\text{TSS}} = \begin{cases}
\mathcal{L}_{\text{FGL}}(\boldsymbol{\mu}, \mathbf{Y}), & p > P(t) \\
\mathcal{L}_{\text{FIBL}}(\boldsymbol{\mu}, \mathbf{Y}). & \text{otherwise}
\end{cases}
\end{equation}
\subsection{Stage II: Detail-Enhanced Diffusion Refinement}

While Stage I generates a coarse radar estimate $\boldsymbol{\mu}$ with proper frequency domain separation, this initial prediction lacks fine-grained precipitation boundaries and detailed intensity variations critical for accurate severe weather characterization. To address this limitation, we introduce a Detail-Enhanced Diffusion Refinement stage that leverages physics-aware frequency-decomposed priors through conditional diffusion modeling.

The conditioning framework integrates multiple information sources to capture complex precipitation dynamics:
\begin{equation}
    \mathbf{C} = [\boldsymbol{\mu}, \mathbf{X}, \tilde{\mathbf{F}}_{\text{LF}}, \tilde{\mathbf{F}}_{\text{HF}}],
\end{equation}
where $\tilde{\mathbf{F}}_{\text{LF}}$ and $\tilde{\mathbf{F}}_{\text{HF}}$ are frequency-decomposed features from our Wavelet-Based Frequency Feature Extractor that provides specialized low-frequency intensity and high-frequency boundary information for targeted refinement. 

The diffusion model then progressively refines precipitation details through iterative denoising:
\begin{equation}
    p_{\theta}(\mathbf{z}_{t-1}|\mathbf{z}_t, \mathbf{C}) = \mathcal{N}(\mathbf{z}_{t-1}; \boldsymbol{\mu}_{\theta}(\mathbf{z}_t, \mathbf{C}), \boldsymbol{\sigma}_{\theta}^2\mathbf{I}).
\end{equation}

\subsubsection{Wavelet-Based Frequency Feature Extractor} 
\begin{figure}[]
\centering
\includegraphics[width=0.45\textwidth]{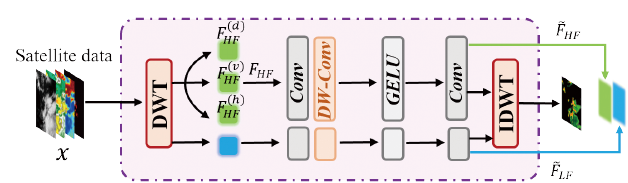}
\caption{Wavelet-Based Frequency Feature Extractor.}
\label{fig:wffe}
\vskip -0.2in
\end{figure}

To provide effective frequency-aware conditioning, we employ a feature extractor that applies DWT to satellite observations $\mathbf{X}$, decomposing them into frequency components $\{{F}_{LF}, {F}^{(d)}_{HF}, {F}^{(v)}_{HF}, {F}^{(h)}_{HF}\}$  (see Fig.~\ref{fig:wffe}). The high-frequency components are aggregated as ${F}_{{HF}} = {F}^{(d)}_{HF} + {F}^{(v)}_{HF} + {F}^{(h)}_{HF}$, and both frequency domains are processed through convolutional transformations:
\begin{equation}
\begin{split}
\tilde{\mathbf{F}}_{\text{LF}} &= \text{Conv}(\text{DWConv}(\text{GELU}(\text{Conv}({F}_{LH})))), \\
\tilde{\mathbf{F}}_{\text{HF}} &= \text{Conv}(\text{DWConv}(\text{GELU}(\text{Conv}({F}_{{HF}})))).
\end{split}
\end{equation}

This extractor enables the diffusion model to leverage both intensity coherence from low-frequency components and boundary precision from high-frequency components for progressive detail enhancement.
\subsubsection{Stage II Loss.}
The second stage optimizes a combined loss function that integrates diffusion denoising with frequency-domain consistency:
\begin{equation}
\mathcal{L}_{\text{Stage II}} = \mathcal{L}_{\text{Diff}} + \lambda_{\text{freq}}\mathcal{L}_{\text{Wavelet}}
\end{equation}
where $\mathcal{L}_{\text{Diff}}=\mathbb{E}_{t,\boldsymbol{\epsilon}}[|\boldsymbol{\epsilon} - \boldsymbol{\epsilon}_{\theta}(\mathbf{z}_t, \mathbf{C}, t)|^2]$ represents the standard diffusion denoising loss that enables progressive refinement through iterative conditioning on $\mathbf{C} = [\boldsymbol{\mu}, \mathbf{X}, \tilde{\mathbf{F}}_{\text{LF}}, \tilde{\mathbf{F}}_{\text{HF}}]$. The frequency consistency term $\mathcal{L}_{\text{Wavelet}}$ ensures proper frequency domain separation in the extracted features $\tilde{\mathbf{F}}_{\text{LF}}$ and $\tilde{\mathbf{F}}_{\text{HF}}$, enabling the diffusion model to leverage physics-aware priors for targeted boundary enhancement while preserving intensity coherence.

\begin{table*}[ht]
\centering
\resizebox{1.80\columnwidth}{!}{
\fontsize{10}{12}\selectfont
\begin{tabular}{c|c|c|cccc|c}
\toprule
\fontsize{10}{12}\selectfont
\multirow{2}{*}{\textbf{Type}} & 
\multirow{2}{*}{\textbf{Model}} & 
\multirow{2}{*}{\textbf{Loss}} & 
\multicolumn{4}{c|}{\textbf{Skill}} & 
\multicolumn{1}{c}{\textbf{Perceptual}}\\
\cline{4-8}
 &  &  & 
\textbf{Avg.CSI}$\uparrow$  & 
\textbf{Avg.HSS}$\uparrow$ & 
\textbf{CSI\_POOL4}$\uparrow$ &
\textbf{CSI\_POOL16}$\uparrow$ &
\textbf{LPIPS}$\downarrow$ 
\\
\midrule
\midrule
\multirow{9}{*}{\centering\shortstack{Pred.}} 
&\multirow{3}{*}{AA-TransUnet~\cite{yang2022aa}} 
& MSE & 0.302 & 0.437 & 0.328 &0.407   & 0.318   \\
& & FACL & 0.313 & 0.452 & 0.365 &0.488   & 0.281  \\
\rowcolor{gray!20}
\cellcolor{white}& \cellcolor{white}&FIBL & 0.314 & \underline{0.455} & 0.366 &0.489   & 0.306 \\
\cline{2-8}

&\multirow{3}{*}{Earthformer~\cite{gao2022earthformer}} 
& MSE & 0.293 & 0.422 & 0.310 &0.376    & 0.360   \\
& & FACL & 0.302 & 0.436 & 0.323 &0.384    & 0.426 \\
\rowcolor{gray!20}
\cellcolor{white}& \cellcolor{white}&FIBL & 0.311 & 0.449 & 0.355 &0.483   & 0.278 \\
\cline{2-8}

&\multirow{3}{*}{Smaat-Unet~\cite{trebing2021smaat}} 
& MSE & 0.293 & 0.421 & 0.309 &0.364   & 0.362   \\
& & FACL & 0.312 & 0.451 & 0.355 &0.440    & 0.375   \\
\rowcolor{gray!20}
\cellcolor{white}& \cellcolor{white}&FIBL & \underline{0.315} & \underline{0.455} & 0.359 &0.448   & 0.361   \\
\cline{1-8}

\multirow{3}{*}{\centering\shortstack{Gen.}} 
& Diffcast~\cite{yu2024diffcast} & -- & 0.310 & 0.448 & \underline{0.391} &0.579   & \underline{0.236} \\
& UVCGAN~\cite{torbunov2023uvcgan}& -- & 0.089 & 0.131 & 0.125 &0.254    & 0.362 \\
& Pix2pix~\cite{akter2024generative}& -- & 0.240 & 0.355 & 0.266 &0.337    & 0.349 \\
& MeanFlow~\cite{geng2025mean}& -- & 0.200 & 0.300 & 0.386 &\underline{0.583}   & 0.300\\
\cline{1-8}
\rowcolor{gray!20}
\cellcolor{white}& Ours {\footnotesize\textcolor{red}{vs AA-TransU\_$\mathcal{L}_{mse}$}} & -- & \textbf{0.327}{\footnotesize\textcolor{red}{+8.2\%}} & \textbf{0.469}{\footnotesize\textcolor{red}{+7.3\%}} & \textbf{0.405}{\footnotesize\textcolor{red}{+23.4\%}} & \textbf{0.592}{\footnotesize\textcolor{red}{+45.4\%}} & \textbf{0.219}{\footnotesize\textcolor{red}{+31.1\%}} \\
\bottomrule
\end{tabular}}
\caption{Quantitative performance comparison on the SEVIR dataset. Gray rows highlight our proposed FIBL loss applied to existing deterministic models and our complete WaveC2R method. The best results are highlighted in bold, and the second-best results are underlined.}
\label{tab:comparison}
\end{table*}
\begin{table*}[t]
\small
\centering
\resizebox{2\columnwidth}{!}{  
\fontsize{10}{12}\selectfont
\setlength{\tabcolsep}{2pt} 
\renewcommand{\arraystretch}{1.2} 
\begin{tabular}{c|c|c|ccccc|ccccc}
\toprule
\fontsize{10}{12}\selectfont
\multirow{2}{*}{\textbf{Type}} & 
\multirow{2}{*}{\textbf{Models}} & \multirow{2}{*}{\textbf{Loss}} & \multicolumn{5}{c}{\textbf{$\uparrow$ CSI}} & \multicolumn{5}{c}{\textbf{$\uparrow$ HSS}} \\
\cline{4-13}
&&  & $\gamma \ge 74$ & $\gamma \ge 133$ & $\gamma \ge 160$ & $\gamma \ge 181$ & $\gamma \ge 219$
& $\gamma \ge 74$ & $\gamma \ge 133$ & $\gamma \ge 160$ & $\gamma \ge 181$ & $\gamma \ge 219$ \\
\midrule
\midrule

\multirow{9}{*}{\centering\shortstack{Pred.}} 
&\multirow{3}{*}{AA-TransUnet~\cite{yang2022aa}} 
&MSE 
 & 0.506 & 0.329& 0.317 & 0.262&0.095 
 & 0.638 & \underline{0.483} &0.476 &0.412& 0.174 \\

&&FACL
&0.512  &\underline{0.330} & 0.317 & 0.270& 0.134
 &0.642  & 0.484 &0.475 & 0.422& 0.235 \\

\rowcolor{gray!20}
\cellcolor{white}&\cellcolor{white}&FIBL
 & 0.503 &0.328 &  0.321&0.276 &0.143 
 &0.635  & 0.482 & \underline{0.479}&0.429 &0.249 \\

\cline{2-13}
  
&\multirow{3}{*}{Earthformer~\cite{gao2022earthformer}}  
&MSE 
 & \underline{0.524} &0.321 &0.304  &0.249 &0.069
 & \textbf{0.653}  &0.474  &0.461 &0.395 &0.129\\

&&FACL
&0.520  &0.302 & 0.288&0.253 & 0.147
& 0.642 &0.441  &0.441 &0.400 & 0.256 \\

\rowcolor{gray!20}
\cellcolor{white}&\cellcolor{white}&FIBL
&0.514 &0.307 &0.315 &0.270 &\underline{0.149} 
&0.644  &0.458&  0.459& 0.422 & \underline{0.260}  \\

\cline{2-13}

&\multirow{3}{*}{Smaat-Unet~\cite{trebing2021smaat}}  
&MSE 
&0.516 &0.324 &0.313  &0.256 & 0.055
 &0.646  & 0.478 &0.471 &0.404 &0.105  \\

&&FACL
&0.508  & 0.319 & 0.317& 0.280 &0.138  
 & 0.637 &0.470& 0.475&0.434 & 0.241  \\

\rowcolor{gray!20}
\cellcolor{white}&\cellcolor{white}&FIBL
  &0.510 &0.314 &0.319  &\underline{0.289} &0.144 
  &0.638  &0.465 &0.477  &0.445  &0.251  \\

\cline{1-13}
\midrule

\multirow{3}{*}{\centering\shortstack{Gen.}} 
&Diffcast~\cite{yu2024diffcast}
&--
 & 0.501 &0.322 &\underline{0.321}  &0.282 &0.126
  &0.629  &0.473 &\underline{0.479} &\underline{0.436} &0.223  \\

&UVCGAN~\cite{torbunov2023uvcgan} 
&--
 & 0.228 &0.094 &0.063  &0.043 & 0.015
 & 0.306 & 0.144 &0.104 & 0.074& 0.028 \\

&Pix2pix~\cite{akter2024generative}
&--
  &0.418  &0.259 & 0.269 & 0.241& 0.014
& 0.547 &0.397  &0.417 &0.384 & 0.028 \\

&MeanFlow~\cite{geng2025mean}
&--
&0.425 & 0.215& 0.177 &0.141 &0.044
&0.550 & 0.333&0.290 & 0.242 &0.084 \\

\cline{1-13}
\rowcolor{gray!20}
& Ours {\footnotesize\textcolor{red}{vs AA-TransU\_$\mathcal{L}_{mse}$}}
& --
& \textbf{0.525}{\footnotesize\textcolor{red}{+3.7\%}} & \textbf{0.337}{\footnotesize\textcolor{red}{+2.4\%}} & \textbf{0.331}{\footnotesize\textcolor{red}{+4.4\%}} & \textbf{0.292}{\footnotesize\textcolor{red}{+11.4\%}} & \textbf{0.152}{\footnotesize\textcolor{red}{+60\%}}
& \underline{0.652} & \textbf{0.490}{\footnotesize\textcolor{red}{+1.4\%}} & \textbf{0.491}{\footnotesize\textcolor{red}{+3.1\%}} & \textbf{0.448}{\footnotesize\textcolor{red}{+8.7\%}} & \textbf{0.263}{\footnotesize\textcolor{red}{+51\%}} \\
\bottomrule
\end{tabular}}
\caption{CSI and HSS performance comparison across different intensity thresholds on SEVIR dataset. Gray rows highlight our proposed FIBL loss applied to existing models and our complete WaveC2R method. The best results are highlighted in bold, and the second-best results are underlined.}
\label{tab:model_comparison}
\vskip -0.2in
\end{table*}
\section{Experiments}
\subsection{Implementation Details}

\noindent\textbf{Dataset.} 
We utilize the SEVIR dataset~\cite{veillette2020sevir}, which integrates multi-source observations including visible data, infrared bands (IR069, IR107), vertically integrated liquid (VIL), and lightning data. The dataset contains over 20,000 storm events from 2017-2019, each with 4-hour temporal coverage at 1 km spatial resolution. We evaluate performance at VIL thresholds of 74, 133, 160, 181, and 219 kg/m². Training details are provided in Appendix B.1.

\noindent\textbf{Evaluation Metrics.} 
To ensure fair comparison with existing methods, we adopt the same evaluation metrics as FACL~\cite{yan2024fourier}. Meteorological performance is assessed using Critical Success Index (CSI) and Heidke Skill Score (HSS) at multiple reflectivity thresholds, which are standard metrics in weather radar evaluation~\cite{fengperceptually}. Image quality evaluation uses Structural Similarity Index (SSIM). For perceptual assessment, we employ Learned Perceptual Image Patch Similarity (LPIPS).

\noindent\textbf{Baseline Comparisons.} 
We compare WaveC2R against CNN-Transformer hybrid models (AA-TransUnet~\cite{yang2022aa}), meteorology-oriented architectures (Earthformer~\cite{gao2022earthformer}), CNN-based networks (Smaat-Unet~\cite{trebing2021smaat}), and generative models including diffusion-based (Diffcast~\cite{yu2024diffcast}), GAN-based methods (UVCGAN~\cite{torbunov2023uvcgan}, Pix2Pix~\cite{akter2024generative}) and Flow-based methods (MeanFlow~\cite{geng2025mean}). MeanFlow details are provided in Appendix A.2.
\begin{figure*}[ht]  
\centering  
\includegraphics[width=0.99\textwidth]{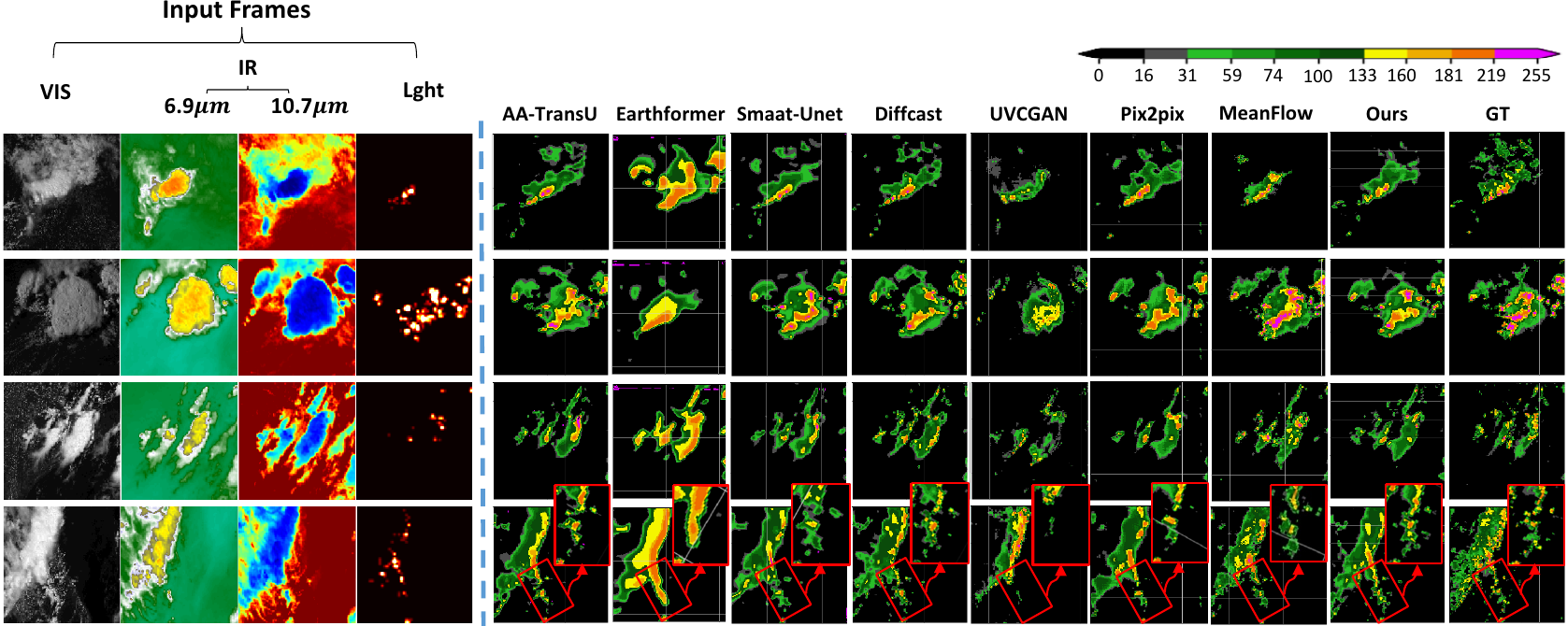}  
\caption{Qualitative comparison of radar reflectivity predictions for four heavy precipitation events from the SEVIR dataset. Red boxes highlight magnified convective regions where our WaveC2R demonstrates superior intensity accuracy and boundary sharpness compared to competing methods. Input modalities include visible, dual-channel infrared, and lightning observations.}
\label{fig:heavy_rain} 
\vskip -0.15in
\end{figure*}

\subsection{Experimental Results}

\noindent\textbf{Metric Comparison.}
Table~\ref{tab:comparison} demonstrates WaveC2R's superiority over state-of-the-art methods. Compared to the best deterministic baseline (AA-TransUnet-MSE), WaveC2R achieves significant improvements: 8.2\% in CSI, 7.3\% in HSS, and 31.1\% in LPIPS, indicating both better skill scores and visual fidelity.

\noindent\textbf{Extreme Weather Performance.}
Table~\ref{tab:model_comparison} presents WaveC2R's strength in severe precipitation detection. At the highest threshold ($\gamma\!\ge\!219$), it achieves 60\% improvement over AA-TransUnet-MSE (0.152 vs 0.095 CSI), while generative models like UVCGAN and Pix2pix fail with near-zero performance (0.015 and 0.014 CSI). This demonstrates WaveC2R's practical value for operational severe weather applications.
 
\noindent\textbf{Qualitative Performance.}
Fig.~\ref{fig:heavy_rain} further illustrates qualitative comparisons for heavy precipitation events. CNN-based models (AA-TransUnet, Smaat-Unet) tend to oversmooth high-reflectivity cores, while Earthformer overestimates moderate-intensity regions. Generative models (Diffcast, UVCGAN, Pix2pix) introduce spurious artifacts. In contrast, WaveC2R accurately preserves both intensity distribution and sharp precipitation boundaries, demonstrating superior reconstruction quality for severe weather systems.

\begin{figure}[]
\centering  
\includegraphics[width=0.47\textwidth]{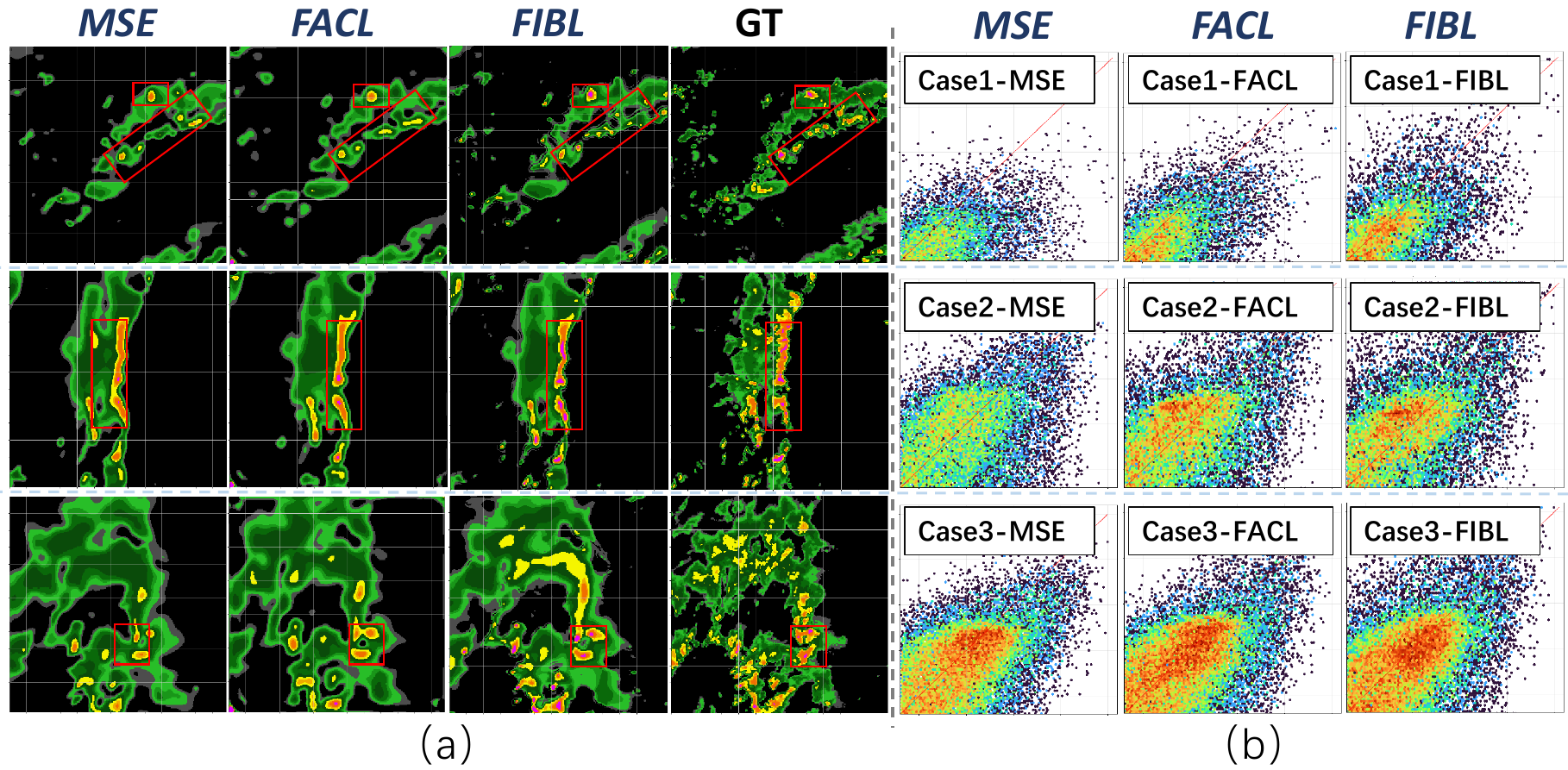}  
\caption{Ablation study of loss functions on Earthformer. (a) Three convective cases shows FIBL achieves superior intensity accuracy and sharper boundaries. (b) Density scatter plots demonstrate better ground truth alignment, particularly in high-intensity regions.}
\label{fig:loss3}
\vskip -0.05in
\end{figure}

\subsection{Ablation Studies}
\begin{table}[h!]
\centering
\resizebox{0.96\columnwidth}{!}{
\begin{tabular}{cccc|c|c}
\hline
WTF & VIS & DEDR&HLF& CSI $\uparrow$ & CSI\_POOL4 $\uparrow$ \\ \hline
\xmark & \cmark & \xmark  & \xmark & 0.311 & 0.328 \\
\cmark & \xmark & \xmark  & \xmark & 0.344 & 0.383 \\
\cmark & \cmark & \xmark & \xmark & 0.360 & 0.413 \\
\cmark & \cmark & \cmark  & \xmark & \underline{0.370} &\underline{0.446}  \\
\hline
\rowcolor{gray!20}
\cmark & \cmark & \cmark & \cmark & \textbf{0.388} & \textbf{0.459} \\ \hline
\end{tabular}}
\caption{Ablation study on the SEVIR dataset based on CSI metric. The best results are highlighted in bold, and the second-best results are underlined.}
\label{tab: ab1}
\vskip -0.25in
\end{table}
\begin{table}[h!]
\centering
\resizebox{0.89\columnwidth}{!}{  
\begin{tabular}{cccc|c|c}
\hline
WTF & VIS & DEDR&HLF& LPIPS $\downarrow$ & SSIM $\uparrow$ \\ \hline
\xmark & \cmark & \xmark  & \xmark & 0.410 & 0.604 \\
\cmark & \xmark & \xmark  & \xmark & 0.331 & 0.597 \\
\cmark & \cmark & \xmark & \xmark & 0.301 & \underline{0.612} \\
\cmark & \cmark & \cmark  & \xmark & \underline{0.227} &0.611  \\
\hline
\rowcolor{gray!20}
\cmark & \cmark & \cmark & \cmark &\textbf{0.219} & \textbf{0.615} \\ \hline
\end{tabular}}
\caption{Ablation study on the SEVIR dataset based on image quality metrics. The best results are highlighted in bold, and the second-best results are underlined.}
\label{tab:ab2}
\vskip -0.25in
\end{table}
To validate the effectiveness of each component, we conduct controlled experiments on the SEVIR dataset. Our method consists of four key components: Wavelet-Temporal-Frequency (WTF) module, visible (VIS) channel, Detail-Enhanced Diffusion Refinement (DEDR), and High-Low Frequency (HLF). Tables~\ref{tab: ab1} and~\ref{tab:ab2} demonstrate that the complete WaveC2R achieves superior performance across all metrics, highlighting the synergistic effects between components for accurate radar retrival. 

\noindent\textbf{Component Analysis.} 
Each component contributes distinctly to reconstruction quality. Removing WTF causes the most significant degradation (CSI: 0.388→0.311, LPIPS: 0.219→0.410), confirming that frequency-aware processing is essential. Excluding VIS, DEDR, and HLF components results in progressive performance drops (CSI: 0.344, 0.360, 0.370 respectively), demonstrating their cumulative importance for accurate radar reflectivity reconstruction. These results validate the effectiveness of each proposed component. Additional ablation studies are provided in Appendix C.

\noindent\textbf{FIBL Loss Analysis.} 
Comparing FIBL against MSE and FACL losses on Earthformer shows FIBL achieves the highest extreme-threshold performance (see Table~\ref{tab:model_comparison}): CSI ($\gamma\!\ge\!219$) improves from 0.069 (MSE) to 0.149 (+116\%), while maintaining superior boundary sharpness without artifacts. Fig.~\ref{fig:loss3} visualizes FIBL's effectiveness in preserving convective cores across diverse cases.

\section*{\large Conclusion}
In this work, we address the challenge of coupled intensity-boundary optimization in satellite-based radar retrieval by introducing WaveC2R, a wavelet-driven coarse-to-refined framework. Based on the insight that precipitation intensity manifests in low-frequency components while boundaries are captured in high-frequency details, WaveC2R decomposes the retrieval task into separate optimization processes through frequency-domain decomposition. Our Wavelet-Temporal-Frequency module enables multi-source feature fusion, while our Frequency-decomposed Intensity-Boundary Loss mitigates energy imbalance in spatial-domain approaches. The diffusion refinement stage leverages physics-aware frequency-decomposed priors for progressive detail enhancement. 

\section*{\large Acknowledgments}
This work was supported by the  Heavy Rainfall Research Foundation of China (No. BYKJ2025M14), Open Project of Key Laboratory of High Impact Weather(special), by the Beijige Foundation of Nanjing Joint Institute for Atmospheric Sciences (NJIAS) under Grant BJG202305, by Key Laboratory of Smart Earth, NO. KF2023YB03-05, and by the National Natural Science Foundation of China (62374031，62331009), and by NSFC-Jiangsu Province (BK20240173).

\renewcommand{\refname}{\large References}
\bibliography{aaai2026}
\end{document}